\documentclass[
 reprint,
 amsmath,amssymb,
 aps,
 superscriptaddress,
 floatfix
]{revtex4-2}

\usepackage{graphicx}
\usepackage{dcolumn}
\usepackage{bm}
\usepackage{booktabs}
\graphicspath{{figure/}{./}}

\begin{document}

\title{QUBO-Based Conformer Selection with Quantum--Classical Hybrid Optimization for Multiconformer Ligand Modeling}

\author{Jiyun Lee}
\affiliation{Department of Chemistry, Yonsei University, Seoul 03722, Republic of Korea}

\author{Haeseong Kim}
\affiliation{Department of Physics, Yonsei University, Seoul 03722, Republic of Korea}

\author{Junho Lee}
\email{awer2072@gmail.com}
\thanks{Co-corresponding author.}
\affiliation{Department of Chemistry, Yonsei University, Seoul 03722, Republic of Korea}

\author{Joonsuk Huh}
\email{ joonsukhuh@gmail.com}
\thanks{Co-corresponding author.}
\affiliation{Department of Chemistry, Yonsei University, Seoul 03722, Republic of Korea}
\affiliation{Department of Quantum Information, Yonsei University, Incheon 21983, Republic of Korea}
\affiliation{Department of Computational Science and Engineering, Yonsei University, Seoul 03722, Republic of Korea}

\date{\today}

\begin{abstract}
A ligand can adopt multiple conformations within the same binding site, and these conformations can interact differently with the surrounding protein. Multiconformer modeling must therefore determine which candidate conformations to retain and what occupancies to assign to them. An established automated approach uses mixed-integer quadratic programming (MIQP) with hard occupancy and conformer-count limits for this selection. Here, we reformulated conformer selection as a quadratic unconstrained binary optimization (QUBO) problem with finite penalties and discrete occupancies. We solved the QUBO using a quantum--classical hybrid solver. We compared this workflow with the MIQP-based qFit-ligand workflow across 83 paired X-ray ligand entries using the same experimental inputs and the same qFit-ligand processing protocol. For each entry, both final models were evaluated against the same composite-omit map generated from the common input structure before the two workflows diverged. Quantum-qFit produced higher ligand-region real-space correlation coefficients (RSCC) in 62 entries (74.7\%) and lower occupancy-weighted torsional strain in 66 entries (79.5\%). Similar directional trends remained in the 28 entries with matched final conformer counts. The molecular Electron Density for Individual Atoms score (EDIAm) decreased in most entries. This result shows that higher regional map agreement did not necessarily correspond to stronger atom-level density support. Overall, the results show that the conformer-selection formulation can change which ligand states are represented from the same experimental data and provide a QUBO framework for studying multiconformer ligand selection.
\end{abstract}


\maketitle

\section{Introduction}

Small molecules bound to proteins do not always adopt a single fixed conformation. A ligand can adopt several conformations within the same binding site~\cite{Perola2004,vanZundert2018}. These conformations can form different interactions with surrounding protein residues~\cite{MobleyDill2009}. Identifying these conformations is therefore important for understanding ligand binding and for structure-based drug design. X-ray crystallography and cryo-electron microscopy (cryo-EM) can provide experimental maps that contain signals from more than one ligand conformation~\cite{Riley2021,Wankowicz2024}. Multiconformer modeling aims to represent these different ligand conformations in a structural model~\cite{vanDenBedem2009,vanZundert2018,Riley2021}.

Building a multiconformer model requires two main steps. First, candidate conformations must be generated. Second, the candidate conformations that are supported by the experimental map must be selected. Methods such as OMEGA~\cite{Hawkins2010}, experimental torsion knowledge distance geometry (ETKDG)~\cite{Riniker2015}, and the Conformer--Rotamer Ensemble Sampling Tool (CREST)~\cite{Grimme2019} can generate plausible ligand conformations. RosettaLigand can also generate ligand binding poses within a protein environment~\cite{Meiler2006}. However, a computationally plausible conformation is not necessarily present in the experimental structure~\cite{McNutt2023,Deller2015}. Candidate conformations must therefore be compared with the experimental map before they are included in the final ligand model.

An existing framework that performs this type of multiconformer modeling is qFit~\cite{vanDenBedem2009,Keedy2015,Riley2021}. qFit automates the construction of multiconformer models using experimental maps, and qFit-ligand applies this approach to small-molecule ligands~\cite{vanZundert2018,Flowers2025}. In qFit-ligand, candidate conformations are first generated. The conformer selection step then determines which candidate conformations are kept and what occupancies are assigned to them. This step is important because later refinement can adjust or remove selected conformers but cannot restore a conformer that was removed during selection. The final ligand model therefore depends not only on which candidate conformations are generated but also on how conformer selection is performed.

In qFit-ligand, this conformer selection problem is solved using mixed-integer quadratic programming (MIQP)~\cite{vanZundert2018,Riley2021,Flowers2025}. MIQP uses continuous variables for conformer occupancies and binary variables for conformer selection. It also uses several hard constraints during selection. The total occupancy cannot exceed one. Each selected conformer must have at least a minimum occupancy. The number of selected conformers cannot exceed a predefined maximum. The total occupancy constraint has a physical meaning. The minimum occupancy and maximum conformer count are empirical limits that help control model complexity. Because these empirical limits are imposed as hard constraints, conformer combinations outside the limits are not considered during selection. Simply relaxing the limits would apply the same change to every ligand. A finite-penalty formulation instead places density fitting and residuals of the encoded selection conditions in the same objective rather than defining all such conditions as absolute feasible boundaries. We hypothesized that this change in formulation would alter which ligand states and occupancies are retained in the final model.

To test this idea, we formulated the conformer selection problem as a quadratic unconstrained binary optimization (QUBO) problem~\cite{Kochenberger2014,Glover2019}. In the QUBO formulation, the selection conditions are represented by penalty terms in the objective function~\cite{Glover2019,Verma2022}. We implemented this formulation within qFit-ligand and named the resulting method Quantum-qFit. Finite penalties are not unique to QUBO and could also be introduced in MIQP. We chose QUBO because the same binary quadratic form can be used by classical Ising-type solvers, quantum annealers, and quantum--classical hybrid solvers~\cite{Lucas2014,Yarkoni2022}. In this study, we used a quantum--classical hybrid solver to test whether the formulation could be executed end to end with a current managed quantum--classical hybrid solver. Quantum-qFit also represents occupancies using discrete values in steps of 0.1. The present comparison therefore evaluates the combined effect of the QUBO selection formulation, occupancy discretization, and the optimization solver.

We then compared Quantum-qFit with qFit-ligand to determine whether the two conformer-selection procedures changed the final ligand models. We used 83 X-ray ligand entries and analyzed four cryo-EM ligand entries as a separate exploratory dataset. Both workflows started from the same experimental input and used the same qFit-ligand settings for map processing, candidate generation, and downstream refinement. For each entry, the composite-omit map was generated once from the common input structure before the workflows diverged. The same map was then used by both branches. Candidate generation and continuous occupancy pre-estimation were run independently, so the intermediate candidate pools were not assumed to be identical. The workflows differed in the conformer-selection formulation, occupancy representation, and optimization solver. We evaluated the final ligand models with three main measures. The real-space correlation coefficient (RSCC) measures agreement between the ligand model and the experimental density over the ligand region~\cite{Jones1991}. Occupancy-weighted torsional strain describes the torsional geometry of the selected ligand conformations~\cite{Rai2019}. The molecular Electron Density for Individual Atoms score (EDIAm) describes atom-level support from the electron density~\cite{Meyder2017}. We also analyzed entries in which both workflows retained the same final number of conformers. This control tested whether the observed differences required a larger final ensemble.

The comparison showed clear differences between the two selection methods. Across the 83 X-ray entries, Quantum-qFit produced higher RSCC in 62 entries (74.7\%) and lower occupancy-weighted torsional strain in 66 entries (79.5\%). The same general pattern remained in the 28 entries for which both methods retained the same final number of conformers. In this subset, RSCC was higher in 21 entries and torsional strain was lower in 20. The observed differences were therefore not limited to cases in which Quantum-qFit produced a larger final ensemble. Protein Data Bank (PDB) entry 7L0K provides an example. Both methods retained two conformers with similar occupancies, but the minor-conformer geometries differed. The density-based measures did not all change in the same direction. EDIAm decreased in 73 of the 83 X-ray entries. The four exploratory cryo-EM entries showed a different pattern. Quantum-qFit had lower RSCC in all four entries and higher EDIAm in three.

Taken together, these results show that conformer selection can influence which ligand states are represented in a multiconformer model. Our study formulates conformer selection guided by experimental maps as a QUBO problem and demonstrates its use within a ligand modeling workflow. The formulation itself is not tied to a particular source of candidate conformations. The runtime behavior of the largest tested instances also motivates controlled evaluation on larger conformer selection problems, although the present timings do not establish a computational speedup. The QUBO formulation therefore provides a basis for studying conformer selection in broader multiconformer modeling problems.

\section{Methods}

\subsection{Selection Formulation and Benchmark Design}

\begin{figure*}[tbp]
    \centering
    \includegraphics[width=\textwidth]{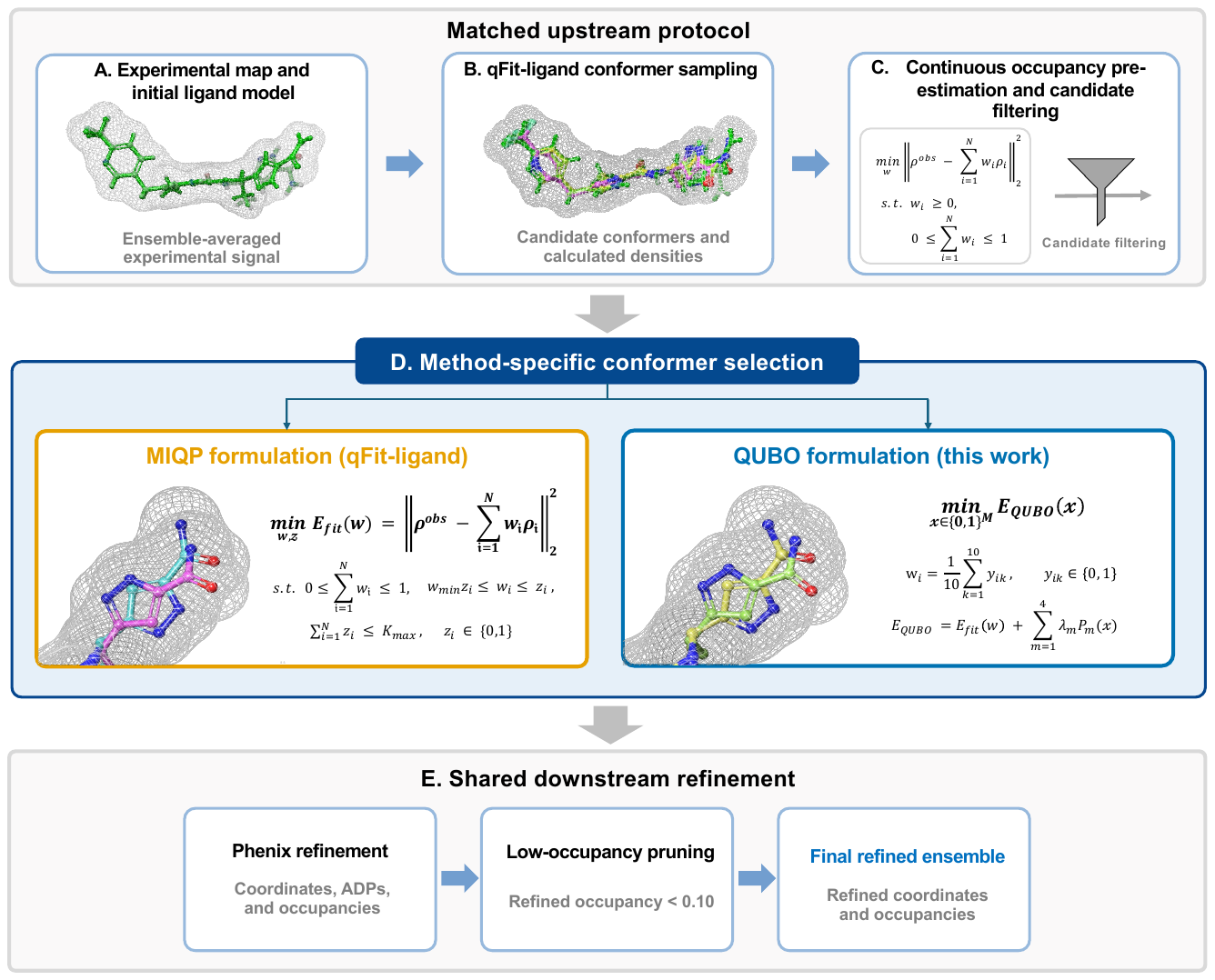}
    \caption{Overview of the two ligand-modeling workflows. Panels A--C show the matched upstream procedure. Both workflows start from the same experimental input and use the same qFit-ligand settings, but candidate generation and occupancy pre-estimation are run independently. Panel D shows the method-specific conformer-selection step. qFit-ligand uses continuous occupancies and hard MIQP constraints. Quantum-qFit uses discrete occupancy bits and finite QUBO penalties and is solved with a quantum--classical hybrid solver. Panel E shows the common downstream procedure. The selected ensembles undergo the same Phenix refinement of coordinates, ADPs, and occupancies, followed by removal of low-occupancy conformers.}
    \label{fig:workflow}
\end{figure*}

We formulated conformer selection against an experimental density map as a quadratic unconstrained binary optimization (QUBO) problem. The objective uses the observed density in the ligand region and the calculated density contribution of each candidate conformer. The formulation does not depend on how the candidate conformations were generated. We evaluated it within qFit-ligand~\cite{vanZundert2018,Riley2021,Wankowicz2024,Flowers2025}, which provides the ligand-sampling framework, the MIQP-based selection procedure, and downstream refinement (Figure~\ref{fig:workflow}). The present benchmark solved the QUBO with a quantum--classical hybrid solver, although the same binary quadratic representation can also be passed to classical binary optimizers.

Throughout this paper, qFit-ligand refers to the MIQP-based workflow. Quantum-qFit refers to the workflow that uses QUBO-based conformer selection. Both workflows used the same experimental input structure and the same composite-omit map generated before the workflows diverged. They also used the same default qFit-ligand settings for map processing, candidate generation, and downstream refinement. Candidate generation and continuous occupancy pre-estimation were run independently, so we do not assume that every intermediate candidate pool was identical. The comparison therefore evaluates the two implemented selection workflows under matched experimental and processing conditions. Quantum-qFit also discretizes occupancies and uses a different optimization solver. The present benchmark does not separate the effects of occupancy discretization, finite-penalty encoding, and solver choice.

\subsection{Benchmark Data and Data Curation}

The primary benchmark was drawn from 135 X-ray ligand entries in the qFit-ligand dataset of Flowers \emph{et al.}~\cite{Flowers2025}. We included the 83 entries for which both workflows produced the outputs required for paired final-model comparisons. Their experimental resolutions ranged from 0.83 to 2.00~\AA. Entry identifiers and paired metrics are recorded in the accompanying benchmark data. The remaining 52 entries did not have complete qFit-ligand outputs for the same paired analysis. Their exclusion may introduce selection bias, so the results should not be generalized automatically to all 135 structures. Each included entry was treated as a paired observation because the two workflows used the same experimental input, the same pre-generated composite-omit map, and the same qFit-ligand processing settings.

Four cryo-electron microscopy (cryo-EM) ligand entries, 8P77, 8P70, 8P6X, and 8P78, were evaluated separately as an exploratory dataset~\cite{Flowers2025}. Their tabulated resolutions were 1.8--2.0~\AA. They were not pooled with the X-ray entries because the imaging and refinement procedures differ.

\subsection{Matched Map Processing and Candidate Generation}

For each entry, a composite-omit map was generated once from the common input structure before the qFit-ligand and Quantum-qFit workflows diverged. The same map was then used by both branches. Candidate conformers were generated independently in the two runs using the same default qFit-ligand sampling protocol~\cite{vanZundert2018,Riley2021,Wankowicz2024,Flowers2025}. The ligand region and calculated conformer densities were constructed using the same qFit-ligand density-fitting procedure and settings in both workflows. Because sampling and filtering were executed independently, the resulting intermediate candidate coordinates and candidate counts were not assumed to be identical entry by entry.

Let $\boldsymbol{\rho}^{\mathrm{obs}}\in\mathbb{R}^{V}$ denote the observed density sampled at $V$ ligand-region voxels and let $\boldsymbol{\rho}_{i}\in\mathbb{R}^{V}$ be the corresponding calculated contribution of candidate conformer $i$. For conformer occupancies $w_i$, the shared density-fitting residual is~\cite{vanDenBedem2009,vanZundert2018,Riley2021,Flowers2025}
\begin{equation}
E_{\mathrm{fit}}(\mathbf{w}) =
\left\|\boldsymbol{\rho}^{\mathrm{obs}}-\sum_{i=1}^{N}w_i\boldsymbol{\rho}_i\right\|_2^2.
\label{eq:fit}
\end{equation}

Expanding Eq.~\ref{eq:fit} gives
\begin{equation}
E_{\mathrm{fit}}(\mathbf{w})=
\mathbf{w}^{\mathsf T}\mathbf{A}\mathbf{w}-2\mathbf{b}^{\mathsf T}\mathbf{w}+C,
\label{eq:fit_matrix}
\end{equation}
where
\begin{equation}
A_{ij}=\langle\boldsymbol{\rho}_i,\boldsymbol{\rho}_j\rangle,\quad
b_i=\langle\boldsymbol{\rho}^{\mathrm{obs}},\boldsymbol{\rho}_i\rangle,\quad
C=\|\boldsymbol{\rho}^{\mathrm{obs}}\|_2^2.
\end{equation}

Each workflow constructed its own $\mathbf{A}$, $\mathbf{b}$, and candidate ordering from the candidates retained in that run using the same procedure.

\subsection{Continuous Occupancy Pre-estimation}

Before discrete conformer selection, each workflow independently estimated continuous occupancies by solving the convex quadratic program (QP) of qFit-ligand~\cite{vanZundert2018,Riley2021,Flowers2025}:
\begin{align}
\min_{\mathbf{w}\in\mathbb{R}^{N}}\quad &E_{\mathrm{fit}}(\mathbf{w}),\\
\mathrm{s.t.}\quad &w_i\geq 0\quad\forall i,\\
&0\leq\sum_{i=1}^{N}w_i\leq 1.
\end{align}

The QP was constructed using CVXPY~\cite{Diamond2016}. The solver backend was not set explicitly in the CVXPY call. Candidate conformers with estimated occupancy below 0.002 were removed before discrete selection. Both workflows used the same threshold and procedure. The pre-estimation and filtering steps were run independently.

\subsection{MIQP Selection in qFit-ligand}

The qFit-ligand MIQP couples continuous occupancies $w_i$ to binary inclusion variables $z_i\in\{0,1\}$~\cite{vanDenBedem2009,vanZundert2018,Riley2021,Flowers2025}:
\begin{align}
\min_{\mathbf{w},\mathbf{z}}\quad &E_{\mathrm{fit}}(\mathbf{w}),\\
\mathrm{s.t.}\quad &w_i\geq 0\quad\forall i,\\
&0\leq\sum_i w_i\leq 1,\\
&w_{\min}z_i\leq w_i\leq z_i\quad\forall i,\\
&\sum_i z_i\leq K_{\max}.
\end{align}

The minimum occupancy was $w_{\min}=0.2$, and the conformer-count limit was $K_{\max}=3$ for X-ray entries and $K_{\max}=2$ for the exploratory cryo-EM entries, following the qFit-ligand implementation~\cite{Riley2021,Flowers2025}. The MIQP was solved locally using SCIP~\cite{Bestuzheva2023}.

\subsection{QUBO Formulation of Conformer Selection}

\paragraph{Binary occupancy encoding.}
For QUBO selection, each pre-refinement occupancy is represented using $L$ binary variables in a unary encoding,
\begin{equation}
w_i=\frac{1}{L}\sum_{k=1}^{L}y_{ik},\qquad y_{ik}\in\{0,1\}.
\label{eq:unary}
\end{equation}

We used $L=10$, which gives occupancy increments of 0.1 and represents $w_{\min}=0.2$ exactly. A binary variable $z_i\in\{0,1\}$ represents whether conformer $i$ is selected within the penalty formulation. The selection stage used these discretized occupancies. The downstream refinement optimized occupancies continuously.

\paragraph{Finite-penalty objective.}
The four selection conditions (total occupancy, occupancy--indicator consistency, minimum selected occupancy, and conformer count) can be written as equalities with nonnegative slack variables and converted to quadratic penalties~\cite{Glover2019,Verma2022}:
\begin{align}
&\sum_i w_i+s_1=1,\\
&w_i+s_{2i}=z_i\quad\forall i,\\
&w_i-s_{3i}=w_{\min}z_i\quad\forall i,\\
&\sum_i z_i+s_4=K_{\max}.
\end{align}

The corresponding QUBO objective is
\begin{equation}
E_{\mathrm{QUBO}}(\mathbf{x})=E_{\mathrm{fit}}(\mathbf{w})+
\sum_{m=1}^{4}\lambda_m P_m(\mathbf{x}),
\label{eq:qubo}
\end{equation}
where
\begin{align}
P_1&=\left(\sum_i w_i+s_1-1\right)^2,\\
P_2&=\sum_i\left(w_i+s_{2i}-z_i\right)^2,\\
P_3&=\sum_i\left(w_i-s_{3i}-w_{\min}z_i\right)^2,\\
P_4&=\left(\sum_i z_i+s_4-K_{\max}\right)^2.
\end{align}

The $P_m$ terms penalize residuals in the four encoded equalities. In our implementation, each occupancy-related slack variable used ten binary bits with weights of 0.1. The conformer-count slack $s_4$ used the same encoding. With $N$ candidate conformers and $L=10$, the complete QUBO contained $M=3NL+N+2L=31N+20$ binary variables. These variables include occupancy bits, occupancy-related slack bits, selection indicators, and conformer-count slack bits.

We used the implementation default $\lambda_1=\lambda_2=\lambda_3=\lambda_4=100$ for all benchmark entries. The penalty weights were not tuned entry by entry. With these weights, an occupancy-related residual of 0.1 contributes 1 to the objective. A residual of 1 in the conformer-count equality contributes 100. Before submission to the hybrid solver, all QUBO coefficients were divided by the positive scale factor $\max(\max_{ij}|Q_{ij}|,1)$. This numerical rescaling does not change the minimizing binary assignment. No independent penalty-sensitivity sweep was performed for the reported benchmark.

The conformer-count slack encoding has an additional limitation. Ten bits with weights of 0.1 allow $s_4$ to range only from 0 to 1. When $K_{\max}=3$, a solution with one active indicator would require $s_4=2$ to satisfy the encoded equality exactly. Our encoding cannot represent this value. Such a solution therefore has a minimum $P_4=1$, which contributes 100 to the objective when $\lambda_4=100$. A nonzero $P_4$ does not by itself show that the original conformer-count inequality was violated. This penalty construction is therefore not an exact encoding of the full MIQP feasible set. We treat this as a limitation and do not assign the observed model differences to a specific penalty term.

After substitution, the QUBO is a binary quadratic objective of the form~\cite{Kochenberger2014,Glover2019}
\begin{equation}
\min_{\mathbf{x}\in\{0,1\}^{M}}\;\mathbf{x}^{\mathsf T}\mathbf{Q}\mathbf{x},
\end{equation}
where $\mathbf{x}$ contains occupancy bits, selection indicators, and slack bits. Here, $M$ is the number of logical binary variables. Our decoder reconstructs occupancies directly from the occupancy bits. It does not independently set an occupancy to zero when the corresponding selection indicator $z_i$ is zero. Because the selection conditions are represented by finite penalties, a returned sample can contain a positive occupancy together with a nonzero indicator-consistency penalty. The archived benchmark outputs do not contain the full returned bit strings. Entrywise $z_i$ values and penalty residuals therefore cannot be reconstructed. Final conformer counts are not used to infer exact satisfaction of every QUBO penalty equality.

\paragraph{Hybrid solver.}
Quantum-qFit constructed a binary quadratic model (BQM) and submitted it through D-Wave \texttt{LeapHybridSampler}~\cite{Yarkoni2022,DWaveHybrid2020}. The lowest-energy returned sample was decoded into pre-refinement occupancies. We did not pass an explicit \texttt{time\_limit} argument. The managed service therefore used its default timing policy. The archived analysis records contain call-to-return wall-clock times but do not include solver-reported QPU access times. The X-ray QUBO instances contained 82--2531 logical binary variables, as measured by $M$. This demonstrates end-to-end execution of the QUBO formulation with a current quantum--classical hybrid solver. It does not isolate a contribution from quantum hardware or establish computational speedup. The archived qFit log reports the development build \texttt{2025.3.dev0+gdf8d1d1c3.d20251104} for the MIQP-based run. The Quantum-qFit branch reports version \texttt{0.1.0}. The archived logs do not record exact version strings for CVXPY, SCIP, or the Ocean SDK.

\subsection{Post-selection Processing and Structural Refinement}

After conformer selection, each workflow underwent the same downstream refinement protocol. For the X-ray entries, the selected multiconformer model was refined with \texttt{phenix.refine} from Phenix 1.21.2-5419~\cite{Adams2010,Liebschner2019,Afonine2012}. Refinement used five macrocycles and optimized individual coordinates, atomic displacement parameters, and occupancies. Coordinate and ADP weights were optimized. Hydrogen atoms were treated with the riding model. After refinement, conformers with refined occupancy below 0.10 were removed. The benchmark workflow did not include an additional Bayesian information criterion model-selection step or a separate occupancy-redistribution step after conformer selection. The exploratory cryo-EM entries followed the corresponding qFit-ligand real-space refinement procedure~\cite{Afonine2018}.

Duplicate pruning within qFit-ligand used a 0.2~\AA{} direct-coordinate criterion during ligand candidate and selection processing. It was not an additional post-refinement Kabsch-aligned filter. The qFit function computes the square root of the mean squared coordinate-component differences without structural alignment. This numerical definition differs by a factor of $\sqrt{3}$ from the conventional atomwise in-place RMSD reported below.

\subsection{Final-Model Evaluation}

All structural metrics were evaluated for the final refined ligand models. The real-space correlation coefficient (RSCC) measures agreement between calculated and observed density over the ligand region~\cite{Jones1991,Deller2015,Smart2018}. For each entry, the composite-omit map was generated once from the common input structure before the two workflows diverged. The same map was used to evaluate both final ligand models. The RSCC analysis used the same qFit validation procedure for both workflows with the 2FOFCWT,PH2FOFCWT map coefficients. The map was scaled over the combined ligand footprint. Both models were then evaluated in the same extracted ligand-region map volume.

Electron Density for Individual Atoms (EDIA) measures local map support for individual atoms. EDIAm summarizes the ligand-level EDIA signal~\cite{Meyder2017,Flowers2025}. Both workflows used the same qFit EDIA implementation, the same composite-omit map, and the same map-processing settings. In the qFit implementation, atom-level EDIA values are first combined within each alternate conformer. The multiconformer ligand value is then formed using the refined conformer occupancies. The crystallographic residuals $R_{\mathrm{work}}$ and $R_{\mathrm{free}}$ assess global model agreement~\cite{Afonine2012}.

Ligand torsional strain was evaluated with the torsion-library approach used in the qFit-ligand benchmark~\cite{Rai2019,Gu2021,Tong2021,Liebeschuetz2021,Flowers2025}. We used the occupancy-weighted model-level value. Throughout this work, \emph{strain} refers to this quantity. Lower strain indicates a lower torsional-energy penalty under this measure. Changes in the ensemble average can arise from changes in conformer geometry, refined occupancy, or both.

Final conformer count denotes the number of ligand conformers remaining after structural refinement and low-occupancy pruning. It is distinct from the number of nonzero selection indicators in the original QUBO sample and from the number of candidates present before the selection-stage duplicate checks.

To characterize occupancy composition, refined occupancies were normalized within each model,
\begin{equation}
p_i=\frac{w_i}{\sum_j w_j},
\end{equation}
and ranked from highest (Rank 1) to lowest. Unoccupied ranks were assigned zero before average rank occupancies were computed across the 83 X-ray entries. This profile describes final ensemble composition. It is not an independent measure of model quality because the available ranks depend on final conformer count.

Paired differences $\Delta$ were defined as Quantum-qFit minus qFit-ligand. Positive $\Delta$ indicates improvement for RSCC and EDIAm. Negative $\Delta$ indicates improvement for strain, $R_{\mathrm{work}}$, and $R_{\mathrm{free}}$. We counted only changes in the favorable direction as improvements. No exact ties occurred for RSCC, strain, or EDIAm among the 83 X-ray entries. Proportions are reported with 95\% Wilson confidence intervals. Two-sided sign tests were used for directional counts, and two-sided Wilcoxon signed-rank tests were used for paired differences. We calculated percentile bootstrap 95\% confidence intervals for median differences from 25,000 entry-level resamples with a fixed random seed. Spearman $r_s$ summarized the association between $\Delta\mathrm{RSCC}$ and $\Delta\mathrm{EDIAm}$. The same descriptive and exploratory tests were applied to the 70-entry and 28-entry control subsets. We did not correct these tests for multiple comparisons.

\subsection{Final Ensemble-Size and Structural Analyses}
\label{sec:ensemble_structural_methods}

The MIQP used a hard $K_{\max}=3$ conformer limit in the X-ray selection stage, whereas Quantum-qFit represented this condition using a finite penalty. We performed two controls on the final models. The first included 70 entries in which Quantum-qFit retained at most three conformers. The second included 28 entries in which both workflows retained the same final number of conformers. These controls test whether the observed differences require a final model exceeding the MIQP limit or containing more conformers than the qFit-ligand model. They do not establish how many conformers were selected before the downstream steps.

For a structural illustration, we selected PDB entry 7L0K, \textit{Plasmodium falciparum} dihydroorotate dehydrogenase bound to DSM784 (ligand XE7, chain A)~\cite{Palmer2021}. This entry belongs to the 28 matched-count cases. Both workflows retained two conformers with nearly identical major and minor occupancies, but the minor-conformer geometries differed. Heavy-atom root-mean-square deviations (RMSDs) were calculated among conformers A and B from the deposited structure, qFit-ligand, and Quantum-qFit. Kabsch-aligned RMSDs~\cite{Kabsch1976} describe differences in conformer shape. We also calculated conventional in-place heavy-atom RMSDs without alignment to describe positional separation within the crystallographic binding-site frame.

Local protein--ligand interactions of the minor conformer were analyzed with the \textit{Show Nonbonded Interactions} tool in Schr\"odinger Maestro~\cite{Maestro2026}. The same settings were applied to the qFit-ligand and Quantum-qFit models. Only residues identified by the software as interacting with the corresponding minor conformer were displayed. We used these contacts to compare local interaction patterns. They were not interpreted as quantitative interaction energies. The archived visualization output does not record the Maestro release string.

\subsection{Runtime Analysis}

Selection-stage wall-clock time was measured from the call to the return of each optimization routine. The qFit-ligand MIQP used a local SCIP solver, whereas Quantum-qFit used a remote managed hybrid solver. The hybrid time can therefore include network transmission, queuing, and managed execution in addition to optimization. We used the number of QUBO logical variables $M$ as a common measure of problem size. Across the X-ray entries, $M$ ranged from 82 to 2531. For each MIQP point, the plotted $M$ value is the QUBO size for the same entry. We report the measured wall-clock times directly and do not fit a scaling law. The archived MIQP environment does not record the CPU model or core count.

\section{Results and Discussion}

\subsection{Benchmark Performance of Quantum-qFit for X-ray Ligand Modeling}

\begin{figure*}[t]
    \centering
    \includegraphics[width=0.96\textwidth]{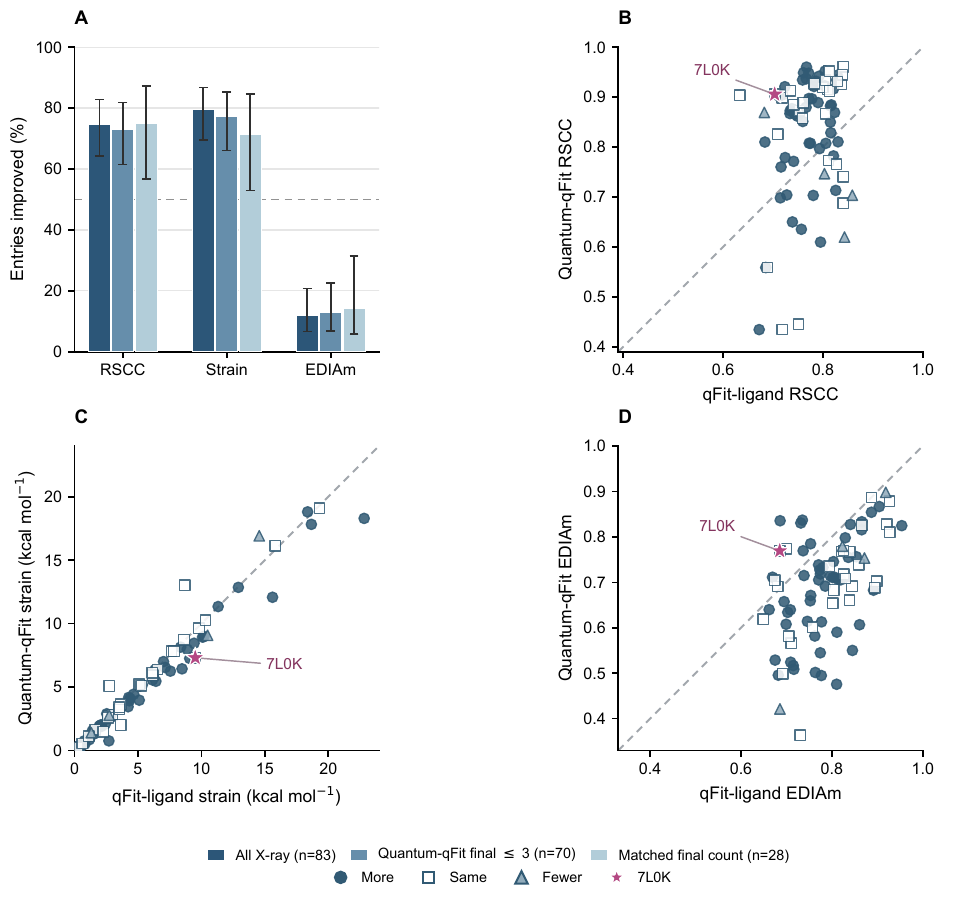}
    \caption{Paired comparison of the 83 X-ray ligand models. (A) Percentage of entries that changed in the favorable direction with Quantum-qFit. Favorable means higher RSCC, lower occupancy-weighted torsional strain, or higher EDIAm. The three bar shades represent all 83 entries, the 70 entries in which Quantum-qFit retained at most three final conformers, and the 28 entries with matched final conformer counts. Error bars show 95\% Wilson confidence intervals. The dashed line marks 50\%. (B--D) Each point represents one paired entry. The horizontal axis shows qFit-ligand and the vertical axis shows Quantum-qFit. The diagonal indicates equal values. Circles mark entries with more final conformers in Quantum-qFit, squares mark equal counts, and triangles mark fewer. The star highlights 7L0K.}
    \label{fig:xray_response}
\end{figure*}

Quantum-qFit produced higher RSCC than qFit-ligand in 62 of the 83 entries (74.7\%). The 95\% Wilson confidence interval (CI) was 64.4--82.8\%. This comparison is shown in Figure~\ref{fig:xray_response}A,B. The two-sided sign-test $p$ value was $7.51\times10^{-6}$. The median paired change in RSCC was $+0.104$ (bootstrap 95\% CI, $+0.067$ to $+0.119$), and the mean change was $+0.058$. RSCC decreased by more than 0.10 in 12 entries, which contributed to the smaller mean than median change.

Quantum-qFit also produced lower occupancy-weighted torsional strain in 66 entries (79.5\%). The 95\% CI was 69.6--86.8\%. This comparison is shown in Figure~\ref{fig:xray_response}A,C. The two-sided sign-test $p$ value was $5.60\times10^{-8}$. The median paired change was $-0.108$~kcal~mol$^{-1}$ (bootstrap 95\% CI, $-0.209$ to $-0.054$~kcal~mol$^{-1}$), and the mean change was $-0.305$~kcal~mol$^{-1}$.

Strain is not part of either selection objective. The lower occupancy-weighted strain therefore reflects the conformers and occupancies that remained after the two workflows rather than direct optimization of strain. RSCC and strain improved together in 49 of 83 entries (59.0\%). This is close to the 49.3 entries expected from their separate improvement frequencies if the two directions were independent. Fisher's exact test gave $p=1.0$. We therefore do not interpret the RSCC and strain improvements as statistically associated. The occupancy-weighted strain value also cannot distinguish a change in conformer geometry from a change in occupancy.

EDIAm changed in the opposite dominant direction. It decreased in 73 of 83 entries and increased in 10. The median $\Delta\mathrm{EDIAm}$ was $-0.092$. The two-sided sign test gave $p=5.80\times10^{-13}$ (Figure~\ref{fig:xray_response}A,D). Higher RSCC coincided with lower EDIAm in 55 entries. Only seven entries improved in all three tabulated metrics. The association between $\Delta\mathrm{RSCC}$ and $\Delta\mathrm{EDIAm}$ was weakly positive (Spearman $r_s=0.23$). This result does not support the idea that the largest RSCC gains were accompanied by the largest EDIAm losses. Whole-structure crystallographic residuals changed little in magnitude. Quantum-qFit had lower $R_{\mathrm{free}}$ in 28 entries, higher $R_{\mathrm{free}}$ in 48, and the same value in seven. The mean paired change was $+0.0001$, and the median was $+0.0002$. The two-sided sign test gave $p=0.029$ (Supporting Information Table S1). The corresponding directional test for $R_{\mathrm{work}}$ was not significant ($p=0.731$). Thus, the ligand-region RSCC trend did not coincide with improved global crystallographic residuals.

RSCC and EDIAm describe different aspects of a ligand model. RSCC measures agreement over the ligand region, whereas EDIAm summarizes support for individual ligand atoms. Across entries, the two metrics were strongly correlated within the qFit-ligand models, with Spearman $r_s=0.95$. The correlation was weaker within the Quantum-qFit models, with $r_s=0.26$ (Supporting Information Figure S3). Both workflows were evaluated against the same composite-omit map generated before they diverged, so this difference is not explained by model-specific maps. Quantum-qFit placed more occupancy on lower-ranked conformers (Figure~\ref{fig:occupancy_profile}). However, the difference in final conformer count was not associated with the EDIAm change across entries ($r_s=-0.03$, $p=0.81$). EDIAm also decreased in 24 of the 28 matched-count entries. These data show that higher ligand-region RSCC does not necessarily imply stronger atom-level density support. They do not identify a single mechanism for the EDIAm decrease.

\subsection{Selection Formulation Alters Final Ensemble Composition}

\begin{figure}[t]
    \centering
    \includegraphics[width=0.98\columnwidth]{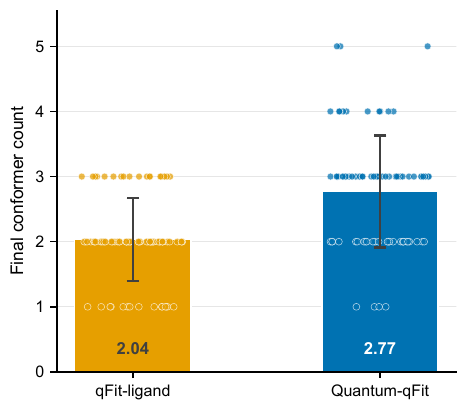}
    \caption{Final ligand conformer counts for the 83 paired X-ray entries. Counts were taken after refinement and low-occupancy pruning. Orange and blue bars show the mean counts for qFit-ligand and Quantum-qFit, respectively. Error bars show standard deviations. Each point represents one entry. Points are jittered horizontally only to make overlapping values visible.}
    \label{fig:conformer_count}
\end{figure}

Quantum-qFit retained more final conformers than qFit-ligand in 51 of the 83 X-ray entries, the same number in 28, and fewer in four (Figure~\ref{fig:conformer_count}). The mean final conformer counts were 2.04 for qFit-ligand and 2.77 for Quantum-qFit. The qFit-ligand models contained one to three conformers, whereas Quantum-qFit models contained one to five. Thirteen Quantum-qFit models ended with four or five conformers, beyond the MIQP selection-stage limit $K_{\max}=3$.

\begin{figure}[t]
    \centering
    \includegraphics[width=0.99\columnwidth]{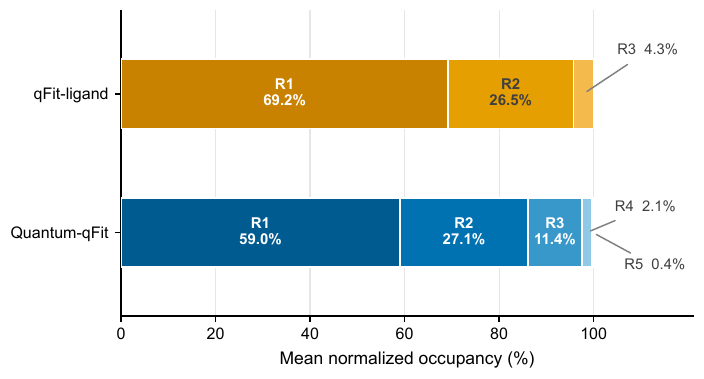}
    \caption{Distribution of refined occupancy across conformer ranks in the 83 X-ray entries. Occupancies were normalized to sum to one within each final model and then ordered from highest to lowest. Rank 1 is therefore the major conformer. Missing ranks were assigned zero before averaging across entries. Orange shades show qFit-ligand and blue shades show Quantum-qFit.}
    \label{fig:occupancy_profile}
\end{figure}

The two workflows also distributed occupancy differently (Figure~\ref{fig:occupancy_profile}). Across the 83 entries, qFit-ligand assigned average normalized occupancies of 69.2\%, 26.5\%, and 4.3\% to Ranks 1--3. Quantum-qFit assigned 59.0\%, 27.1\%, 11.4\%, 2.1\%, and 0.4\% to Ranks 1--5. Quantum-qFit therefore placed about 10 percentage points less occupancy on the major conformer. Most of this difference appeared in Ranks 3--5. The occupancy profile describes how the final ensembles differ. It is not an independent measure of model quality because the available ranks depend on final conformer count.

The larger Quantum-qFit ensembles cannot be attributed to the conformer-count penalty alone. Quantum-qFit also discretizes occupancies, represents selection conditions with finite penalties, and uses a different solver. The penalties have different numerical effects for residuals of similar scale. With $\lambda_m=100$, an occupancy-related residual of 0.1 contributes 1 to the objective. A residual of 1 in the conformer-count equality contributes 100. Among the 51 entries with more Quantum-qFit final conformers, 38 still contained no more than three conformers. These final counts were also allowed by the MIQP limit. Identifying the mechanism would require selection-stage occupancies, indicator values, and penalty residuals. Final conformer counts alone are not sufficient.

The predominant X-ray pattern remained when models with more than three final conformers were excluded (Figure~\ref{fig:xray_response}A). This control left 70 entries. RSCC increased in 51 entries (72.9\%, 95\% CI, 61.5--81.9\%). Strain decreased in 54 entries (77.1\%, 95\% CI, 66.0--85.4\%). The median changes were $+0.104$ for RSCC and $-0.081$~kcal~mol$^{-1}$ for strain. EDIAm decreased in 61 of the 70 entries. The predominant RSCC and strain directions therefore did not require a final Quantum-qFit model with more than three conformers.

Differences also remained when the two workflows produced the same final conformer count. This matched-count subset contained 28 entries. RSCC increased in 21 entries (75.0\%, 95\% CI, 56.6--87.3\%). Strain decreased in 20 entries (71.4\%, 95\% CI, 52.9--84.7\%). The median changes were $+0.111$ for RSCC and $-0.043$~kcal~mol$^{-1}$ for strain. EDIAm decreased in 24 entries. The mean strain change was $+0.035$~kcal~mol$^{-1}$ despite the negative median because a few entries had larger strain increases. The Wilcoxon signed-rank test for strain gave $p=0.077$. The directional sign test gave $p=0.036$, but this exploratory test was not adjusted for multiple comparisons.

Higher RSCC occurred in 74.7\% of all X-ray entries, 72.9\% of the at-most-three-conformer subset, and 75.0\% of the matched-count subset. A larger final ensemble was therefore not required for the predominant RSCC difference. The 13 entries with four or five final Quantum-qFit conformers also showed higher RSCC in 11 cases and lower strain in 12, while EDIAm decreased in 12. These final-model controls show that the two conformer selection methods can produce different ensembles even when final conformer count is controlled. They do not identify which QUBO condition caused the difference, and they do not establish a contribution from the hybrid solver.

\subsection{Structural Consequences of Alternative Conformer Selection in 7L0K}

PDB entry 7L0K illustrates how the two workflows can produce different ligand models even when the final conformer count is the same. The structure is \textit{Plasmodium falciparum} dihydroorotate dehydrogenase bound to inhibitor DSM784 (ligand XE7, chain A)~\cite{Palmer2021}. Both workflows retained two conformers with similar occupancies. qFit-ligand assigned occupancies of 0.79 and 0.21 to conformers A and B. Quantum-qFit assigned 0.81 and 0.19 (Table~\ref{tab:7l0k_metrics}). The RSCC values were 0.703 and 0.906, and the EDIAm values were 0.685 and 0.769. The occupancy-weighted torsional strain values were 9.52 and 7.31~kcal~mol$^{-1}$ for qFit-ligand and Quantum-qFit, respectively. The increase in EDIAm is opposite to the predominant direction across the X-ray benchmark. Because the two models had the same conformer count and nearly the same occupancies, we compared the geometries and binding-site positions of the retained conformers.

\begin{table}[t]
    \centering
    \small
    \setlength{\tabcolsep}{4pt}
    \caption{Final refined model statistics for the illustrative X-ray entry 7L0K. $\Delta$ is calculated as Quantum-qFit minus qFit-ligand. For RSCC and EDIAm, a positive $\Delta$ means that the Quantum-qFit value is higher. For torsional strain, $R_{\mathrm{work}}$, and $R_{\mathrm{free}}$, a negative $\Delta$ means that the Quantum-qFit value is lower. Strain is the occupancy-weighted model-level value used throughout the benchmark.}
    \label{tab:7l0k_metrics}
    \resizebox{\columnwidth}{!}{%
    \begin{tabular}{lccc}
        \toprule
        Metric & qFit-ligand & Quantum-qFit & $\Delta$ \\
        \midrule
        Final conformer count & 2 & 2 & 0 \\
        Occupancy, major conformer (A) & 0.79 & 0.81 & $+0.02$ \\
        Occupancy, minor conformer (B) & 0.21 & 0.19 & $-0.02$ \\
        RSCC & 0.703 & 0.906 & $+0.203$ \\
        EDIAm & 0.685 & 0.769 & $+0.084$ \\
        Torsional strain (kcal~mol$^{-1}$) & 9.52 & 7.31 & $-2.21$ \\
        $R_{\mathrm{work}}$ & 0.2006 & 0.1968 & $-0.0038$ \\
        $R_{\mathrm{free}}$ & 0.2313 & 0.2315 & $+0.0002$ \\
        \bottomrule
    \end{tabular}%
    }
\end{table}

\begin{figure*}[t]
    \centering
    \includegraphics[width=\textwidth]{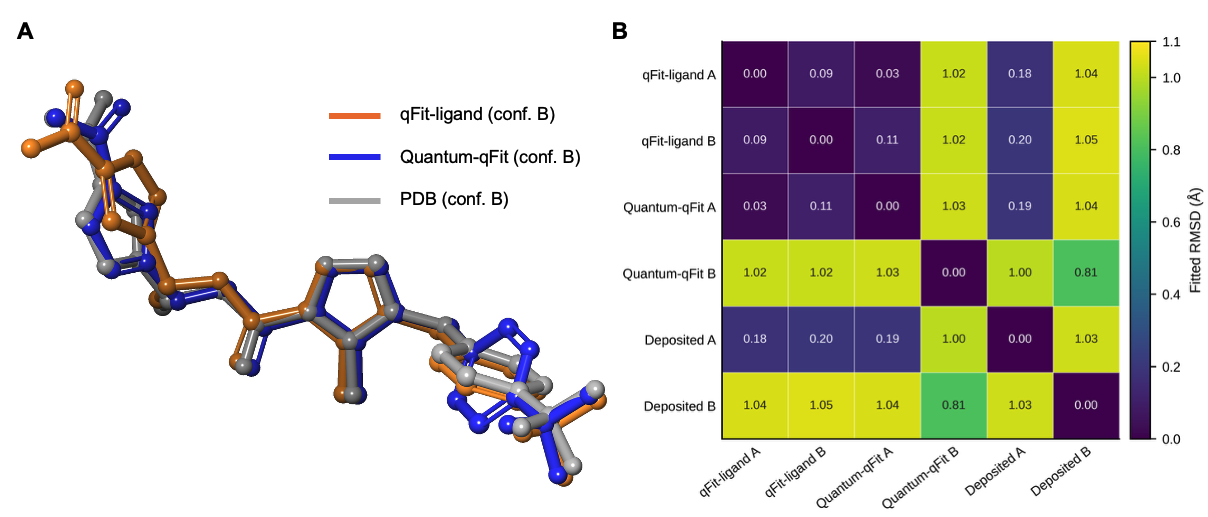}
    \caption{Structural comparison of the refined 7L0K ligand models in chain A. (A) Overlay of minor conformer B from qFit-ligand (orange), Quantum-qFit (blue), and the deposited structure (gray) after structural alignment. (B) Pairwise heavy-atom RMSDs after Kabsch alignment for conformers A and B from the three models. Smaller values indicate more similar ligand shapes after alignment. Because this fitting removes overall translation and rotation, these RMSDs do not measure how far the conformers are separated in the binding site. The corresponding A--B RMSDs calculated without alignment are discussed in the Results.}
    \label{fig:case_7l0k_rmsd}
\end{figure*}

Pairwise fitted heavy-atom RMSDs were calculated among conformers A and B from qFit-ligand, Quantum-qFit, and the deposited structure (Figure~\ref{fig:case_7l0k_rmsd}B). The major conformers were highly similar. qFit-ligand A and Quantum-qFit A differed by only 0.03~\AA. Their fitted RMSDs to deposited conformer A were 0.18 and 0.19~\AA, respectively. The larger difference appeared in conformer B. After Kabsch alignment, the qFit-ligand minor conformer differed from its own major conformer by only 0.09~\AA. Its fitted RMSDs to deposited conformers A and B were 0.20 and 1.05~\AA. In contrast, the Quantum-qFit minor conformer differed from its major conformer by 1.03~\AA. Its fitted RMSDs to deposited conformers A and B were 1.00 and 0.81~\AA, respectively.

The in-place comparison explains why the small fitted RMSD of the qFit-ligand pair does not mean that the two conformers occupy the same binding-site position. Without alignment, the qFit-ligand A--B heavy-atom RMSD was approximately 0.48~\AA. The corresponding Quantum-qFit value was approximately 1.05~\AA. The qFit duplicate check uses an unaligned coordinate-component RMSD rather than a Kabsch-fitted RMSD. In the 7L0K selection-stage output, the qFit-ligand A--B value by this internal definition was approximately 0.27~\AA, which is above the 0.2~\AA\ duplicate threshold. The retained qFit-ligand conformers can therefore have nearly identical fitted shapes while remaining distinct under the coordinate-space duplicate criterion.

\begin{figure*}[t]
    \centering
    \includegraphics[width=0.96\textwidth]{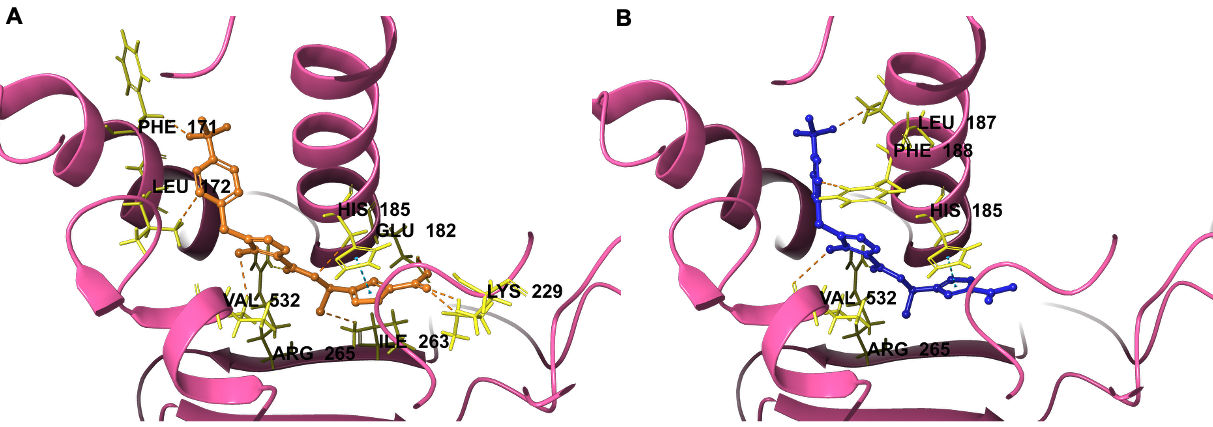}
    \caption{Local protein--ligand contacts for minor conformer B in the refined 7L0K models. (A) qFit-ligand with the ligand shown in orange. (B) Quantum-qFit with the ligand shown in blue. Contacts were identified with the \textit{Show Nonbonded Interactions} tool in Schr\"odinger Maestro using the same settings for both models. Dashed lines indicate geometrical nonbonded contacts reported by the software. Residue labels show the protein residues associated with those contacts in each model. Protein-residue colors are display colors only and do not represent interaction strength. The panels are therefore used to compare contact patterns qualitatively rather than interaction energies.}
    \label{fig:case_7l0k_map_interactions}
\end{figure*}

We next assessed whether the difference in minor-conformer geometry was associated with a different local protein--ligand interaction pattern. The \textit{Show Nonbonded Interactions} analysis in Schr\"odinger Maestro was applied to the minor conformer from each refined model using the same settings. Only residues identified by the software as interacting with the corresponding minor conformer were displayed. The qFit-ligand minor conformer showed contacts involving Phe171, Leu172, and Lys229, whereas the Quantum-qFit minor conformer showed contacts involving Leu187 and Phe188. His185, Val532, and Arg265 were identified in both models (Figure~\ref{fig:case_7l0k_map_interactions}). The different residue sets therefore reflect different software-identified local interaction patterns rather than different manual choices of displayed residues.

Thus, 7L0K shows how conformer selection can affect structural interpretation even when the final conformer count and occupancies are similar. The two workflows retained nearly identical major conformers but different minor-conformer geometries. These different minor conformers were also associated with different local protein--ligand interaction patterns. This example therefore illustrates that conformer selection can influence not only how many ligand states are represented, but also which minor state is represented and how its local protein environment is interpreted.

\subsection{Exploratory Evaluation of Cryo-EM Ligand Models}

\begin{figure*}[tbp]
    \centering
    \includegraphics[width=0.93\textwidth]{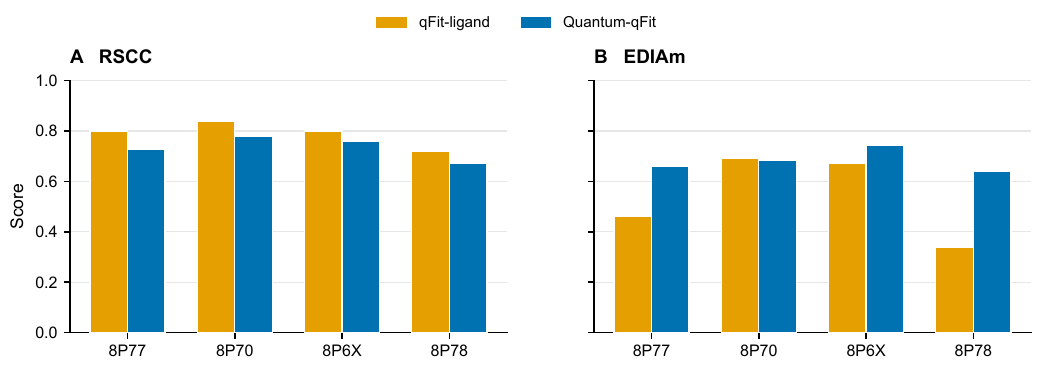}
    \caption{Exploratory comparison of four cryo-EM ligand entries. (A) RSCC and (B) EDIAm for qFit-ligand (orange) and Quantum-qFit (blue). Higher values indicate greater agreement with the map for RSCC and stronger atom-level density support for EDIAm. The same PDB entry order is used in both panels.}
    \label{fig:cryo_em}
\end{figure*}

Quantum-qFit produced lower RSCC than qFit-ligand in all four exploratory cryo-EM entries, 8P77, 8P70, 8P6X, and 8P78 (Figure~\ref{fig:cryo_em}A). The paired differences ranged from $-0.043$ to $-0.071$, with a mean of $-0.055$. This direction is opposite to the mean X-ray RSCC change of $+0.058$. EDIAm increased in three of the four cryo-EM entries, with a mean $\Delta\mathrm{EDIAm}$ of $+0.141$ (Figure~\ref{fig:cryo_em}B). It increased from 0.340 to 0.640 for 8P78, from 0.460 to 0.658 for 8P77, and from 0.670 to 0.742 for 8P6X. For 8P70, EDIAm changed from 0.690 to 0.685. Because this dataset contains only four entries and differs from the X-ray set in map processing, we treat the cryo-EM analysis as exploratory.

The cryo-EM pattern therefore differed from the dominant X-ray pattern. The present data do not show whether this difference arises from map processing, selection settings, refinement, or another factor. Four entries are also too few to establish a general cryo-EM performance trend.

\subsection{Runtime Behavior across Conformer-Selection Problem Sizes}

\begin{figure}[tbp]
    \centering
    \includegraphics[width=0.99\columnwidth]{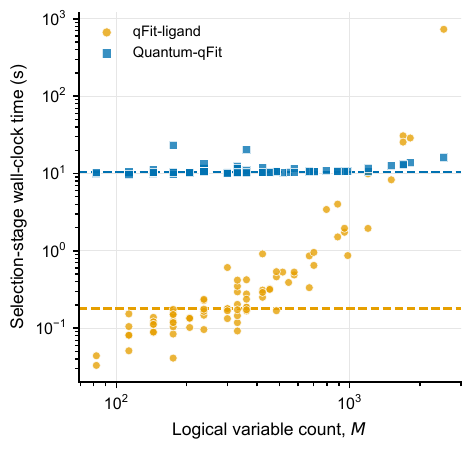}
    \caption{Selection-stage wall-clock time for the 83 X-ray entries as a function of QUBO problem size $M$. Each point represents one entry. Orange circles show qFit-ligand MIQP times from the local SCIP solver. Blue squares show Quantum-qFit times from the remote D-Wave hybrid solver. The horizontal position of each MIQP point uses the QUBO size for the same entry so that both methods can be displayed on one problem-size axis. Both axes are logarithmic, and dashed lines show the median wall-clock times. Because the two methods ran in different execution environments, the plot shows observed timing behavior only and does not demonstrate a computational speedup.}
    \label{fig:runtime}
\end{figure}

The local MIQP solver was faster for most of the tested ligand-selection problems, while the largest instances showed a different timing pattern (Figure~\ref{fig:runtime}). The median wall-clock time was 0.179~s for qFit-ligand and 10.473~s for the remote Quantum-qFit hybrid workflow. The tested QUBOs ranged from $M=82$ to $2531$ logical variables. For the four largest instances ($M\geq1694$), the local MIQP times ranged from 25.478 to 732.261~s. The remote hybrid times ranged from 12.750 to 16.090~s and varied less across these entries. Quantum-qFit did not pass an explicit \texttt{time\_limit} argument to \texttt{LeapHybridSampler}. The managed solver therefore used its default timing policy. We measured call-to-return wall-clock time and did not archive solver-reported QPU access time. The methods ran in different computing environments, and solver optimality was not controlled across them. These timings therefore do not demonstrate intrinsic scaling behavior or computational speedup from QUBO or quantum hardware. They motivate a controlled timing study on larger conformer-selection problems.

\section{Conclusion and Outlook}

We formulated ligand conformer selection guided by experimental maps as a quadratic unconstrained binary optimization problem and implemented it within the qFit-ligand modeling framework as Quantum-qFit. Both workflows used the same experimental inputs, the same composite-omit map generated before the workflows diverged, and the same default qFit-ligand processing settings. Candidate generation and occupancy pre-estimation were executed independently, so the comparison does not assume identical intermediate candidate pools. Quantum-qFit also changed occupancy representation, finite-penalty encoding, and optimization solver together, and the present study therefore does not separate the contribution of these components or of quantum hardware.

Across 83 X-ray ligand entries, Quantum-qFit produced higher ligand-region RSCC in 62 entries (74.7\%) and lower occupancy-weighted torsional strain in 66 (79.5\%). These predominant directions remained when Quantum-qFit retained no more than three final conformers and when both workflows retained the same final conformer count. Because both final models were evaluated against the same composite-omit map for each entry, the RSCC difference is not explained by the use of model-specific maps. PDB entry 7L0K further showed that two models with the same number of conformers and similar occupancies can retain different minor-conformer geometries and binding-site positions. Together, these results show that conformer selection can affect which ligand states are represented, not only how many states are retained.

The results also define the current limits of interpretation. EDIAm decreased in 73 of 83 X-ray entries, so higher ligand-region RSCC did not correspond to stronger atom-level density support under this measure. The four exploratory cryo-EM entries showed a different pattern from the X-ray benchmark, and 52 of the original 135 X-ray entries lacked complete paired outputs. The occupancy-weighted strain value cannot separate conformer geometry from changes in refined occupancies. In addition, the conformer-count slack encoding does not reproduce the full MIQP feasible range exactly, and the archived QUBO outputs do not retain the full bit strings needed to reconstruct entrywise penalty residuals. The observed model differences therefore should not be assigned to a specific penalty, discretization choice, or solver.

The hybrid solver was slower for most of the present ligand problems, although its wall-clock times varied less across the largest tested instances. This observation does not establish a speedup. More broadly, the QUBO formulation provides a concrete alternative representation of density-guided conformer selection that can be tested with different penalty encodings, occupancy representations, and optimizers. Controlled studies on larger ligand ensembles are a natural next step.

\section*{Data and Software Availability}

The experimental structures used in this study are publicly available through the Protein Data Bank. Machine-readable benchmark data are provided as Supporting Information CSV files. The Quantum-qFit source code is provided as a ZIP archive with the Supporting Information.

\section*{Supporting Information}

Supporting Information includes a PDF with additional details of the benchmark and QUBO encoding, paired statistical summaries, final conformer-count distributions, EDIAm analyses, and exploratory cryo-EM results. It also includes entry-level X-ray and cryo-EM benchmark data in CSV format and the Quantum-qFit source code as a ZIP archive.

\section*{Author Information}

\textbf{Corresponding Authors}\\
Junho Lee -- Email: awer2072@gmail.com\\
Joonsuk Huh -- Email: joonsukhuh@yonsei.ac.kr

\section*{Author Contributions}

Jiyun Lee developed the Quantum-qFit workflow, performed the analyses, and drafted the manuscript. Haeseong Kim and Junho Lee contributed to method development, structural analysis, and manuscript review. Joonsuk Huh conceived and supervised the study, contributed to interpretation, and revised the manuscript. All authors approved the final manuscript.

\section*{Notes}

The authors declare no competing financial interest.

\section*{Acknowledgments}

This work was supported by the National Research Foundation of Korea funded by the Ministry of Science and ICT (RS-2023-NR068116 and RS-2025-03532992), the Korea Health Industry Development Institute funded by the Ministry of Health and Welfare through the Korean ARPA-H Project (RS-2025-25456722), the Institute for Information \& Communications Technology Promotion grant funded by the Korean government (MSIP) (No. 2019-0-00003), and the Yonsei University Research Fund (2025-22-0140).

During manuscript preparation, the authors used ChatGPT (OpenAI) to assist with language editing and manuscript organization. All scientific content, analyses, interpretations, and final wording were reviewed and approved by the authors, who take full responsibility for the manuscript.


\bibliography{reference}

@article{Keedy2015,
  author = {Keedy, Daniel A. and Fraser, James S. and van den Bedem, Henry},
  title = {Exposing Hidden Alternative Backbone Conformations in X-ray Crystallography Using {qFit}},
  journal = {PLoS Computational Biology},
  year = {2015},
  volume = {11},
  number = {10},
  pages = {e1004507},
  doi = {10.1371/journal.pcbi.1004507}
}

@article{vanZundert2018,
  author = {van Zundert, Gydo C. P. and Hudson, Brandi M. and de Oliveira, Saulo H. P. and Keedy, Daniel A. and Fonseca, Rasmus and Heliou, Amelie and Suresh, Pooja and Borrelli, Kenneth and Day, Tyler and Fraser, James S. and van den Bedem, Henry},
  title = {{qFit}-ligand Reveals Widespread Conformational Heterogeneity of Drug-Like Molecules in X-Ray Electron Density Maps},
  journal = {Journal of Medicinal Chemistry},
  year = {2018},
  volume = {61},
  number = {24},
  pages = {11183--11198},
  doi = {10.1021/acs.jmedchem.8b01292}
}

@article{Flowers2025,
  author = {Flowers, Jessica and Echols, Nathaniel and Correy, Galen J. and Jaishankar, Priyadarshini and Togo, Takaya and Renslo, Adam R. and van den Bedem, Henry and Fraser, James S. and Wankowicz, Stephanie A.},
  title = {Expanding Automated Multiconformer Ligand Modeling to Macrocycles and Fragments},
  journal = {eLife},
  year = {2025},
  volume = {14},
  pages = {RP103797},
  doi = {10.7554/eLife.103797.3}
}

@article{Jones1991,
  author = {Jones, T. A. and Zou, J.-Y. and Cowan, S. W. and Kjeldgaard, M.},
  title = {Improved Methods for Building Protein Models in Electron Density Maps and the Location of Errors in These Models},
  journal = {Acta Crystallographica Section A},
  year = {1991},
  volume = {47},
  number = {2},
  pages = {110--119},
  doi = {10.1107/S0108767390010224}
}

@article{Meyder2017,
  author = {Meyder, Agnes and Nittinger, Eva and Lange, Gudrun and Klein, Robert and Rarey, Matthias},
  title = {Estimating Electron Density Support for Individual Atoms and Molecular Fragments in X-ray Structures},
  journal = {Journal of Chemical Information and Modeling},
  year = {2017},
  volume = {57},
  number = {10},
  pages = {2437--2447},
  doi = {10.1021/acs.jcim.7b00391}
}

@article{Gu2021,
  author = {Gu, Shuo and Smith, Matthew S. and Yang, Ying and Irwin, John J. and Shoichet, Brian K.},
  title = {Ligand Strain Energy in Large Library Docking},
  journal = {Journal of Chemical Information and Modeling},
  year = {2021},
  volume = {61},
  number = {9},
  pages = {4331--4341},
  doi = {10.1021/acs.jcim.1c00368}
}

@article{Kochenberger2014,
  author = {Kochenberger, Gary and Hao, Jin-Kao and Glover, Fred and Lewis, Mark and L{\"u}, Zhipeng and Wang, Haibo and Wang, Yang},
  title = {The Unconstrained Binary Quadratic Programming Problem: A Survey},
  journal = {Journal of Combinatorial Optimization},
  year = {2014},
  volume = {28},
  number = {1},
  pages = {58--81},
  doi = {10.1007/s10878-014-9734-0}
}

@article{Lucas2014,
  author = {Lucas, Andrew},
  title = {Ising Formulations of Many NP Problems},
  journal = {Frontiers in Physics},
  year = {2014},
  volume = {2},
  pages = {5},
  doi = {10.3389/fphy.2014.00005}
}

@article{Yarkoni2022,
  author = {Yarkoni, Sheir and Raponi, Elena and B{\"a}ck, Thomas and Schmitt, Sebastian},
  title = {Quantum Annealing for Industry Applications: Introduction and Review},
  journal = {Reports on Progress in Physics},
  year = {2022},
  volume = {85},
  number = {10},
  pages = {104001},
  doi = {10.1088/1361-6633/ac8c54}
}

@misc{DWaveHybrid2020,
  author = {{D-Wave Quantum Inc.}},
  title = {Sampling with Hybrid Solvers: {D-Wave Leap} Hybrid Solver Service},
  howpublished = {D-Wave Quantum Computing Products Documentation},
  year = {2020},
  url = {https://docs.dwavequantum.com/en/latest/industrial_optimization/index_hybrid.html},
  note = {Documentation consulted October 16, 2026}
}

@article{Diamond2016,
  author = {Diamond, Steven and Boyd, Stephen},
  title = {{CVXPY}: A Python-Embedded Modeling Language for Convex Optimization},
  journal = {Journal of Machine Learning Research},
  year = {2016},
  volume = {17},
  number = {83},
  pages = {1--5}
}

@article{Bestuzheva2023,
  author = {Bestuzheva, Ksenia and Besan{\c{c}}on, Mathieu and Chen, Wei-Kun and Chmiela, Antonia and Donkiewicz, Tim and van Doornmalen, Jasper and Eifler, Leon and Gaul, Oliver and Gamrath, Gerald and Gleixner, Ambros and others},
  title = {Enabling Research through the {SCIP} Optimization Suite 8.0},
  journal = {ACM Transactions on Mathematical Software},
  year = {2023},
  volume = {49},
  number = {2},
  pages = {1--21},
  doi = {10.1145/3585516}
}

@article{Liebschner2019,
  author = {Liebschner, Dorothee and Afonine, Pavel V. and Baker, Matthew L. and Bunk{\'o}czi, G{\'a}bor and Chen, Vincent B. and Croll, Tristan I. and Hintze, Bradley and Hung, Li-Wei and Jain, Swati and McCoy, Airlie J. and others},
  title = {Macromolecular Structure Determination Using X-rays, Neutrons and Electrons: Recent Developments in Phenix},
  journal = {Acta Crystallographica Section D: Structural Biology},
  year = {2019},
  volume = {75},
  number = {10},
  pages = {861--877},
  doi = {10.1107/S2059798319011471}
}

@article{Afonine2012,
  author = {Afonine, Pavel V. and Grosse-Kunstleve, Ralf W. and Echols, Nathaniel and Headd, Jeffrey J. and Moriarty, Nigel W. and Mustyakimov, Marat and Terwilliger, Thomas C. and Urzhumtsev, Alexandre and Zwart, Peter H. and Adams, Paul D.},
  title = {Towards Automated Crystallographic Structure Refinement with phenix.refine},
  journal = {Acta Crystallographica Section D: Biological Crystallography},
  year = {2012},
  volume = {68},
  number = {4},
  pages = {352--367},
  doi = {10.1107/S0907444912001308}
}

@article{Afonine2018,
  author = {Afonine, Pavel V. and Poon, Billy K. and Read, Randy J. and Sobolev, Oleg V. and Terwilliger, Thomas C. and Urzhumtsev, Alexandre and Adams, Paul D.},
  title = {Real-Space Refinement in {PHENIX} for Cryo-EM and Crystallography},
  journal = {Acta Crystallographica Section D: Structural Biology},
  year = {2018},
  volume = {74},
  number = {6},
  pages = {531--544},
  doi = {10.1107/S2059798318006551}
}

@article{Kabsch1976,
  author = {Kabsch, Wolfgang},
  title = {A Solution for the Best Rotation to Relate Two Sets of Vectors},
  journal = {Acta Crystallographica Section A},
  year = {1976},
  volume = {32},
  number = {5},
  pages = {922--923},
  doi = {10.1107/S0567739476001873}
}

@article{vanDenBedem2009,
  author = {van den Bedem, Henry and Dhanik, Ankur and Latombe, Jean-Claude and Deacon, Ashley M.},
  title = {Modeling Discrete Heterogeneity in X-ray Diffraction Data by Fitting Multi-Conformers},
  journal = {Acta Crystallographica Section D: Biological Crystallography},
  year = {2009},
  volume = {65},
  number = {10},
  pages = {1107--1117},
  doi = {10.1107/S0907444909030613}
}

@article{Riley2021,
  author = {Riley, Blake T. and Wankowicz, Stephanie A. and de Oliveira, Saulo H. P. and van Zundert, Gydo C. P. and Hogan, Daniel W. and Fraser, James S. and Keedy, Daniel A. and van den Bedem, Henry},
  title = {{qFit} 3: Protein and Ligand Multiconformer Modeling for X-Ray Crystallographic and Single-Particle Cryo-{EM} Density Maps},
  journal = {Protein Science},
  year = {2021},
  volume = {30},
  number = {1},
  pages = {270--285},
  doi = {10.1002/pro.4001}
}

@article{Wankowicz2024,
  author = {Wankowicz, Stephanie A. and Ravikumar, Ashraya and Sharma, Shivani and Riley, Blake T. and Raju, Akshay and Hogan, Daniel W. and Flowers, Jessica and van den Bedem, Henry and Keedy, Daniel A. and Fraser, James S.},
  title = {Automated Multiconformer Model Building for X-Ray Crystallography and Cryo-{EM}},
  journal = {eLife},
  year = {2024},
  volume = {12},
pages = {RP90606},
doi = {10.7554/eLife.90606.3}
}

@article{Rai2019,
  author = {Rai, Brajesh K. and Sresht, Vishnu and Yang, Qingyi and Unwalla, Raymond J. and Tu, Meihua and Mathiowetz, Alan M. and Bakken, Gregory A.},
  title = {Comprehensive Assessment of Torsional Strain in Crystal Structures of Small Molecules and Protein--Ligand Complexes Using Ab Initio Calculations},
  journal = {Journal of Chemical Information and Modeling},
  year = {2019},
  volume = {59},
  number = {10},
  pages = {4195--4208},
  doi = {10.1021/acs.jcim.9b00373}
}

@article{Liebeschuetz2021,
  author = {Liebeschuetz, John W.},
  title = {The Good, the Bad, and the Twisted Revisited: An Analysis of Ligand Geometry in Highly Resolved Protein--Ligand X-Ray Structures},
  journal = {Journal of Medicinal Chemistry},
  year = {2021},
  volume = {64},
  number = {11},
  pages = {7533--7543},
  doi = {10.1021/acs.jmedchem.1c00228}
}

@article{McNutt2023,
  author = {McNutt, Andrew T. and Bisiriyu, Fatimah and Song, Sophia and Vyas, Ananya and Hutchison, Geoffrey R. and Koes, David Ryan},
  title = {Conformer Generation for Structure-Based Drug Design: How Many and How Good?},
  journal = {Journal of Chemical Information and Modeling},
  year = {2023},
  volume = {63},
  number = {21},
  pages = {6598--6607},
  doi = {10.1021/acs.jcim.3c01245}
}

@article{Tong2021,
  author = {Tong, Jiahui and Zhao, Suwen},
  title = {Large-Scale Analysis of Bioactive Ligand Conformational Strain Energy by Ab Initio Calculation},
  journal = {Journal of Chemical Information and Modeling},
  year = {2021},
  volume = {61},
  number = {3},
  pages = {1180--1192},
  doi = {10.1021/acs.jcim.0c01197}
}

@article{Verma2022,
  author = {Verma, Amit and Lewis, Mark W.},
  title = {Penalty and Partitioning Techniques to Improve Performance of QUBO Solvers},
  journal = {Discrete Optimization},
  year = {2022},
  volume = {44},
  pages = {100594},
  doi = {10.1016/j.disopt.2020.100594}
}

@article{MobleyDill2009,
  author = {Mobley, David L. and Dill, Ken A.},
  title = {Binding of Small-Molecule Ligands to Proteins: ``What You See'' Is Not Always ``What You Get''},
  journal = {Structure},
  year = {2009},
  volume = {17},
  number = {4},
  pages = {489--498},
  doi = {10.1016/j.str.2009.02.010}
}

@article{Perola2004,
  author = {Perola, Emanuele and Charifson, Paul S.},
  title = {Conformational Analysis of Drug-Like Molecules Bound to Proteins: An Extensive Study of Ligand Reorganization upon Binding},
  journal = {Journal of Medicinal Chemistry},
  year = {2004},
  volume = {47},
  number = {10},
  pages = {2499--2510},
  doi = {10.1021/jm030563w}
}

@article{Deller2015,
  author = {Deller, Marc C. and Rupp, Bernhard},
  title = {Models of Protein-Ligand Crystal Structures: Trust, but Verify},
  journal = {Journal of Computer-Aided Molecular Design},
  year = {2015},
  volume = {29},
  number = {9},
  pages = {817--836},
  doi = {10.1007/s10822-015-9833-8}
}

@article{Smart2018,
  author = {Smart, Oliver S. and Horsk{\'y}, Vladim{\'i}r and Gore, Swanand and Svobodov{\'a} Va{\v r}ekov{\'a}, Radka and Bendov{\'a}, Veronika and Kleywegt, Gerard J. and Velankar, Sameer},
  title = {Validation of Ligands in Macromolecular Structures Determined by X-Ray Crystallography},
  journal = {Acta Crystallographica Section D: Structural Biology},
  year = {2018},
  volume = {74},
  pages = {228--236},
  doi = {10.1107/S2059798318002541}
}

@article{Glover2019,
  author = {Glover, Fred and Kochenberger, Gary and Du, Yu},
  title = {A Tutorial on Formulating and Using {QUBO} Models},
  journal = {4OR},
  year = {2019},
  volume = {17},
  number = {4},
  pages = {335--371},
  doi = {10.1007/s10288-019-00424-y}
}

@article{Adams2010,
  author = {Adams, Paul D. and Afonine, Pavel V. and Bunk{\'o}czi, G{\'a}bor and Chen, Vincent B. and Davis, Ian W. and Echols, Nathaniel and Headd, Jeffrey J. and Hung, Li-Wei and Kapral, Gary J. and Grosse-Kunstleve, Ralf W. and McCoy, Airlie J. and Moriarty, Nigel W. and Oeffner, Robert and Read, Randy J. and Richardson, David C. and Richardson, Jane S. and Terwilliger, Thomas C. and Zwart, Peter H.},
  title = {{PHENIX}: A Comprehensive Python-Based System for Macromolecular Structure Solution},
  journal = {Acta Crystallographica Section D: Biological Crystallography},
  year = {2010},
  volume = {66},
  number = {2},
  pages = {213--221},
  doi = {10.1107/S0907444909052925}
}

@misc{Maestro2026,
  author = {{Schrödinger, LLC}},
  title = {Maestro},
  year = {2026},
  address = {New York, NY}
}

@article{Hawkins2010,
  author  = {Hawkins, Paul C. D. and Skillman, A. Geoffrey and Warren, Gregory L. and Ellingson, Benjamin A. and Stahl, Matthew T.},
  title   = {Conformer Generation with OMEGA: Algorithm and Validation Using High Quality Structures from the Protein Databank and Cambridge Structural Database},
  journal = {Journal of Chemical Information and Modeling},
  year    = {2010},
  volume  = {50},
  number  = {4},
  pages   = {572--584},
  doi     = {10.1021/ci100031x}
}

@article{Riniker2015,
  author  = {Riniker, Sereina and Landrum, Gregory A.},
  title   = {Better Informed Distance Geometry: Using What We Know To Improve Conformation Generation},
  journal = {Journal of Chemical Information and Modeling},
  year    = {2015},
  volume  = {55},
  number  = {12},
  pages   = {2562--2574},
  doi     = {10.1021/acs.jcim.5b00654}
}

@article{Grimme2019,
  author  = {Grimme, Stefan},
  title   = {Exploration of Chemical Compound, Conformer, and Reaction Space with Meta-Dynamics Simulations Based on Tight-Binding Quantum Chemical Calculations},
  journal = {Journal of Chemical Theory and Computation},
  year    = {2019},
  volume  = {15},
  number  = {5},
  pages   = {2847--2862},
  doi     = {10.1021/acs.jctc.9b00143}
}

@article{Meiler2006,
  author  = {Meiler, Jens and Baker, David},
  title   = {ROSETTALIGAND: Protein-Small Molecule Docking with Full Side-Chain Flexibility},
  journal = {Proteins},
  year    = {2006},
  volume  = {65},
  number  = {3},
  pages   = {538--548},
  doi     = {10.1002/prot.21086}
}

@article{Palmer2021,
  author  = {Palmer, Michael J. and others},
  title   = {Potent Antimalarials with Development Potential Identified by Structure-Guided Computational Optimization of a Pyrrole-Based Dihydroorotate Dehydrogenase Inhibitor Series},
  journal = {Journal of Medicinal Chemistry},
  year    = {2021},
  volume  = {64},
  number  = {9},
  pages   = {6085--6136},
  doi     = {10.1021/acs.jmedchem.1c00173}
}

\end{document}